\documentclass[twocolumn,prl,showpacs,amsmath,amssymb,superscriptaddress,floatfix]{revtex4}
\usepackage{graphicx,color} 

\newcommand{\be}{\begin{eqnarray}}
\newcommand{\ee}{\end{eqnarray}}

\DeclareMathOperator{\csch}{csch}

\DeclareMathOperator{\erf}{erf}
\DeclareMathOperator{\erfi}{erfi}
\DeclareMathOperator{\erfc}{erfc}

\def\1{\mathit 1}

\def\infint{\int_{-\infty}^{\infty}}

\def\Dth{D_{\text{th}}^+}
\def\DdS{D_{\text{dS}}^+}

\def\bra#1{\left\langle {#1} \right\rvert}
\def\ket#1{\left\lvert {#1} \right\rangle}

\def\outprod#1#2{\ket {#1}\!\bra {#2}}

\def\avg#1{\left\langle {#1} \right\rangle}
\def\abs#1{\left\lvert{#1}\right\rvert}

\begin{document}

\title{Entangling Power of an Expanding Universe}

\author{Greg Ver Steeg}
\email{gregv@caltech.edu}
\affiliation{California Institute of Technology, Pasadena, California 91125, USA}

\author{Nicolas C. Menicucci}
\email{nmen@princeton.edu}
\affiliation{Department of Physics, Princeton University, Princeton, New Jersey 08544, USA}
\affiliation{Department of Physics, The University of Queensland, Brisbane, Queensland 4072, Australia}

\date{\today}

\begin{abstract}
We show that entanglement can be used to detect spacetime curvature.  Quantum fields in the Minkowski vacuum are entangled with respect to local field modes.  This entanglement can be swapped to spatially separated quantum systems using standard local couplings.  A single, inertial field detector in the exponentially expanding (de~Sitter) vacuum responds as if it were bathed in thermal radiation in a Minkowski universe.  We show that using two inertial detectors, interactions with the field in the thermal case will entangle certain detector pairs that would not become entangled in the corresponding de~Sitter case.  The two universes can thus be distinguished by their entangling power.
\end{abstract}


\maketitle


Information in curved spacetime has played a prominent role in the attempt to understand the interface between quantum physics and gravity~\cite{Bekenstein1973, Hawking1975, Gibbons1977, Bousso2002a}.  While abstract properties of curved-space quantum fields (including their entanglement) can be studied directly~\cite{Hawking2001,Goheer2003,Bousso2002,Summers1987,Ball2006}, an operational approach involving observers with detectors historically has been a critical component of theoretical progress in this area~\cite{Gibbons1977,Birrell1982}.  With the birth of quantum information theory~\cite{Nielsen2000}, quantum systems could now be analyzed in terms of their use for information-theoretic tasks like quantum computation~\cite{Nielsen2000}, quantum teleportation~\cite{Bennett1993}, and quantum cryptography~\cite{Lo2007}.  Entanglement is a phenomenon that is uniquely quantum mechanical in nature~\cite{Masanes2007} and can be considered both an information-theoretic and a physical resource~\cite{Blume-Kohout2002}.  It is known that the Minkowski vacuum possesses long-range entanglement~\cite{Summers1987} that can be swapped to local inertial systems using standard quantum coupling mechanisms~\cite{Reznik2005}.  Variations on this theme can be considered, including accelerating detectors~\cite{Massar2006}, thermal states~\cite{Braun2005}, and curved spacetime.  Our focus will be on curvature.  For this, we choose an exponentially expanding (de~Sitter) universe~\cite{Bousso2002, Spradlin2001} for its simplicity and because of its importance to cosmology~\cite{Guth1997}.

We wish to demonstrate a connection between a physical property of spacetime (curvature) and an information-theoretic resource (entanglement).  While it is possible to directly study the entanglement present in a quantum field in de~Sitter spacetime, this sometimes leads to difficulties~\cite{Goheer2003} that are not present in a more operational approach.  Still, it is known that entanglement between field modes can directly encode a spacetime's curvature parameters~\cite{Ball2006}.  Motivated by a desire to be as operational as possible, we examine how curvature affects a field's usefulness as an {\it entangling resource}---i.e., its ability to entangle distant quantum systems (``detectors'') using purely local interactions.  We begin by reviewing the response of a single, inertial detector interacting with a massless, conformally coupled scalar field. The result in the vacuum de Sitter case is identical to that in the case of a thermal ensemble of field particles in flat spacetime~\cite{Birrell1982,Gibbons1977}.  Next, we ask the question, {\em can entanglement be used to distinguish de~Sitter vacuum expansion from Minkowski-space heating?}  We show that with two detectors on comoving trajectories, there exists a parameter regime in which the local systems that couple to the field will become entangled despite the presence of extra thermal noise in each individual detector.  Interestingly, this region of parameter-space in the expanding case is a {\it proper subset} of the same region in the locally equivalent thermal case.  Thus, while both universes affect a local inertial detector in exactly the same way, entanglement between two detectors can be used to distinguish them.


We start with the following experimental setup, which is nearly identical to that used by Reznik {\it et al.\/}~\cite{Reznik2005}, using units where $\hbar = c = k_B = 1$.  We pose our problem completely in operational terms, but our goal is to show proof of principle---not necessarily practicality of the method.  We suppose that the inhabitants of a particular planet launch a satellite into space to measure the temperature of the universe they inhabit.  On board this satellite is a qubit (a two-level quantum system), initially in the ground state $\ket 0$, that gets coupled locally and for a limited time to a scalar field using a simple De~Witt monopole coupling~\cite{DeWitt1979}.  The time-dependent interaction Hamiltonian for this detector is, in the interaction picture,
\begin{align}
\label{eq:Hint}
	H_I(\tau) = \eta(\tau) \phi\bigl(x(\tau)\bigr) \bigl(e^{+i \Omega \tau} \sigma^+ + e^{-i \Omega \tau} \sigma^- \bigr)\;,
\end{align}
where $\tau$ is the proper time of the satellite, $\eta(\tau)$ is a weak time-dependent coupling parameter (which we'll call the detector's ``window function''), $x(\tau)$ is the worldline of the satellite, $\phi(x)$ is the field operator at the spacetime location~$x$, and the rest represents the interaction-picture Pauli operator~$\sigma_x(\tau)$ for the local qubit with (tunable) energy gap~$\Omega$.  Roughly speaking, the detector works by inducing oscillations between the two levels at a strength governed by the local value of the field.

From now on, we refer to this qubit as a ``detector,'' although the process of ``detection'' includes only the field interaction (before projective measurement).  We wish to examine when two such detectors become entangled through their local interactions with the field, so we delay classical readout to allow for general quantum postprocessing, which may be necessary to show violation of a Bell inequality~\cite{Masanes2006}.

The window function $\eta(\tau)$ is used to turn the detector on and off, but the transitions must be sufficiently smooth so as not to excite the field too much in the process~\cite{Sriramkumar1996}.  Beyond this requirement, on physical grounds, our results should not depend on the details of the window function as long as it is approximately time bounded, so we will always choose $\eta(\tau)$ to be proportional to a Gaussian, $\eta(\tau) = \eta_0 e^{- (\tau - \tau_0)^2 / 2\sigma^2}$, where $\eta_0 = \eta(\tau_0) \ll 1$ is a small unitless constant that enforces the weak-coupling limit and allows us to use perturbation theory.  This window function approximates the detector being ``on'' when $\abs {\tau- \tau_0} \lesssim \sigma$ and ``off'' the rest of the time and also has a nice analytic form.

Without loss of generality, we can set $\tau_0 = 0$.  To lowest nontrivial order in $\eta_0$, the qubit after the interaction (but before readout) will be found in the state $\rho = A \outprod 1 1 + (1-A) \outprod 0 0$, where
\begin{align}
\label{eq:Adef}
	A = \infint d\tau \infint d\tau' \, \eta(\tau) \eta(\tau') e^{-i \Omega (\tau-\tau')} D^+ \bigl( x(\tau);x(\tau') \bigr)\;,
\end{align}
where $D^+(x;x') = \avg{ \phi(x) \phi(x') }$ is the Wightman function for the field, with expectation taken with respect to the state of the field (assumed to be a zero-mean Gaussian state, but not necessarily the vacuum).  Repeated measurement in the $\{\ket 0, \ket 1\}$ basis for a variety of values of $\Omega$ allows for determination of the state of the detector as a function of $\Omega$~\footnote{More general measurements will be required to demonstrate entanglement between two such detectors, though.}.  As is clear from Eq.~\eqref{eq:Adef}, the state is completely determined by the detector response function~$D^+ \bigl( x(\tau);x(\tau') \bigr)$, which is the Wightman function taken at two different proper times along the worldline of the detector~\cite{Birrell1982}.

We consider two possible universes.  The first is Minkowski, $ds^2 = dt^2 - \sum_{i=1}^3 dx_i^2$, with the field in a thermal state with temperature~$T$ with respect to the inertial trajectory $\{x_i\}=\text{(constant)}$.  The second is a de~Sitter universe, $ds^2 = dt^2 - e^{2\kappa t} \sum_{i=1}^3 dx_i^2$, where $\kappa$ is the expansion rate, in the conformal vacuum.  
The conformal vacuum is the natural choice in this case because it is the unique, coordinate-independent vacuum state dictated by the symmetries of the spacetime.  Furthermore, it can be justified on physical grounds because the conformal vacuum coincides with the massless limit of the adiabatic vacuum for de~Sitter space~\cite{Birrell1982}.  Thus, we can think of this analysis as applying to the following two ways of adiabatically modifying the Minkowski vacuum: (1) very slowly heating the universe to a temperature~$T$, and (2) very slowly ramping up the de~Sitter expansion rate (from zero) to a final value of~$\kappa$.

The variables $\{x_i\}$ are comoving coordinates, and $t$ is cosmic time.  (Since the Minkowski metric is the special case $\kappa=0$, this terminology carries over to it, as well.)  In both universes, worldlines of constant $\{x_i\}$ are inertial trajectories (geodesics), and intervals of proper time equal those of cosmic time ($\Delta \tau = \Delta t$).  In both cases, the scalar field~$\phi(x)$ is massless and conformally coupled~\cite{Birrell1982}, satisfying $[\square_x + \tfrac 1 6 R(x)] \phi(x) = 0$, where the Ricci scalar~$R(x) = 12 \kappa^2$ is a constant proportional to the expansion rate~$\kappa$. 

Gibbons and Hawking~\cite{Gibbons1977} showed that the detector response function for any inertial observer in the de~Sitter case is {\it exactly the same} as that of a detector at rest in a thermal bath of field particles with temperature~$T = \kappa / 2 \pi$ in flat spacetime.  Thus, a single detector alone cannot distinguish between the two cases if it forever remains on a given inertial trajectory.  In both cases considered above, the detector is at rest in the comoving frame and thus,
\begin{align}
\label{eq:Ddetect}
	D_T^+ \bigl( x(\tau);x(\tau') \bigr) = -\frac {T^2} {4} \csch^2 [\pi  T (t - t' - i\epsilon)]\;,
\end{align}
where the subscript $T$ indicates that this is a detector response function for a thermal state at temperature~$T$. When the satellite begins sending back measurement data, the reconstructed $A(\Omega)$ is found to be consistent with the detector being at rest in a thermal bath of field particles at a small but nonzero temperature $T$.  If the inhabitants wish to know whether this perceived thermality is a result of heating or expansion, though, they must be more creative.

Obviously, they could use astrophysical clues (like we have done on Earth) and/or Doppler-shift measurements~\footnote{The thermal Minkowski case exhibits Doppler shifting for detectors at different velocities~\cite{Birrell1982}; the vacuum de~Sitter case does not~\cite{Gibbons1977}.} to determine whether their universe is expanding or not, but we are going to restrict them to using only satellite-mounted detectors of the sort described above on fixed inertial trajectories.  If the detectors are to be useful, then, they will need more than one.

We propose the following alternative that makes use of entanglement to distinguish the two universes.  We imagine two satellites, each having many qubits that interact locally with the scalar field.  (Having many detectors allows access to many copies of the same state.)  We assume that the satellites have no initial entanglement with each other and that the qubits each begin in the ground state.  After interacting with the field, measurement is delayed to allow for general quantum operations (local to each satellite) on the multitude of qubits on board.  In the end, however, the only data that can be transmitted back to the home planet are measurement results, plus information about the postprocessing and the particular measurements performed.

In an attempt to be as simple as possible, we analyze the case of two inertial detectors, $a$ and $b$, on the comoving trajectories $x_1 = \pm L/ 2$ (with $x_2=x_3=0$).  Due to the homogeneity and isotropy of space in both scenarios, this case is remarkably general---but not entirely so since one could imagine the detectors in motion with respect to each other (beyond the relative motion generated by any expansion).  For simplicity, we'll also require that the two detectors have synchronized local clocks with $\tau_{a,b} = t$, equal resonant frequencies $\Omega_{a,b} = \Omega$, and identical window functions $\eta_{a,b}(\tau) = \eta_0 e^{- \tau^2 / 2\sigma^2}$.  Finally, we desire that $L \gg \sigma$ so that the detector-field interactions can be considered noncausal events~\footnote{Although the Gaussian window functions technically have tails that extend forever, none of the results change if we assume a smooth cutoff of the Gaussian (to zero) around, say, $10 \sigma$ as long as both $L$ and $T^{-1}$ are still much larger than this.}.  As we shall see, these restrictions will still allow the inhabitants, located at $x_i=0$, to distinguish expansion from heating.

By spatial symmetry, each detector alone must respond using the detector response function from Eq.~\eqref{eq:Ddetect} and thus provides no useful information.  The only hope, then, is in the correlations between the detectors.  We will focus on those correlations that signal the presence of {\it entanglement} of the detectors after interaction with the field.  For a pair of qubits, the negativity~\cite{Vidal2002} of a state is nonzero if and only if the systems are entangled~\cite{Horodecki1996}.  Since we have access to (by assumption) multiple copies of an entangled state of pairs of qubits, a local measurement protocol (on the many copies of the state) always exists to verify entanglement by showing a violation of a Bell inequality~\cite{Masanes2006,Horodecki1997}.  This can be verified by a third party using classical data received from both satellites.

We will focus on finding the regimes in which entanglement is nonzero, rather than on the magnitude of the entanglement for two reasons.  First, the amount of extractable entanglement is small enough to be impractical as a resource and will depend on the details of the detector coupling.  Second, we are primarily interested in understanding a qualitative difference between the quantum behavior of curved and flat spacetime; examining entanglement ensures that this is a genuinely quantum mechanical effect~\cite{Masanes2007}.

An analogous calculation to Reznik's~\cite{Reznik2005} shows that the negativity of the joint state of the qubits is $N = \max \bigl( \abs X - A, 0 \bigr)$, where $A$ is the individual detector response from Eq.~\eqref{eq:Adef}, while $X$ is defined as
\begin{align}
\label{eq:Xdef}
	X &=  -\infint dt \int_{-\infty}^t dt'\, \eta(t) \eta(t') e^{i \Omega (t+t')} \nonumber \\
		& \qquad\times \bigl[D^+\bigl(x_a(t);x_b(t') \bigr) + D^+\bigl(x_b(t);x_a(t') \bigr) \bigr] \nonumber \\
	&= -2\int_{t'<t} dt\, dt'\, \eta(t) \eta(t') e^{i \Omega (t+t')} D^+\bigl(x_a(t);x_b(t') \bigr)  \;.
\end{align}
The limits of integration enforce time ordering~\cite{Peskin1995}, so we can use the Wightman function as shown.  This is useful because symmetry of the two detectors means that $D^+\bigl(x_a(t);x_b(t') \bigr) = D^+\bigl(x_b(t);x_a(t') \bigr)$, a fact used to obtain the second line.  This integral measures the {\it amplitude} that the detectors will exchange a virtual particle, while $A$ measures the {\it probability} that each detector becomes excited either by absorbing or emitting a particle.

\begin{figure}[htbp]
\begin{center}
\includegraphics[width= 8.5 cm,angle=0]{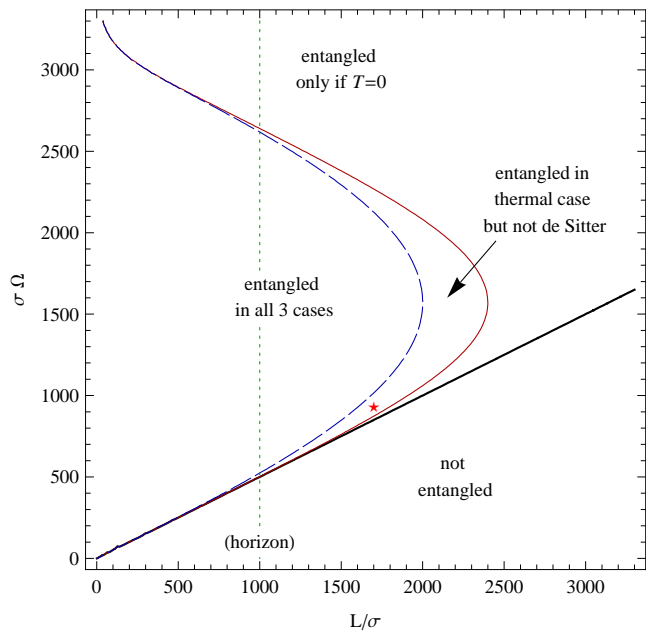}
\caption{Entanglement profile for detector pairs in several universes---$\sigma$ is detection time, $\Omega$ is detector resonance frequency, $L$ is detector separation.  The slanted black line is the entanglement cutoff in the Minkowski vacuum case (entangled above, separable below).  The solid red curve is the thermal Minkowski cutoff, and the dashed blue curve is the de~Sitter vacuum cutoff, both with perceived local temperatures satisfying $2 \pi T = 10^{-3} \sigma^{-1}$.  The de~Sitter horizon distance ($10^3 \sigma$) is given by the dotted green line.  The red star indicates one particular detector setup that could be used to distinguish expansion from heating.}
\label{fig:threshold}
\end{center}
\end{figure}

We begin by considering when the qubits become entangled when $T=0$.  (We also {\it define} $\kappa \equiv 2 \pi T$ from now on so we can talk about expansion rates in terms of the associated Gibbons-Hawking temperature.) This case corresponds to the one considered by Reznik~\cite{Reznik2005} using different window functions.  In the $T=0$ case, the Wightman function used in $X$ is
\begin{align}
\label{eq:DMink}
	D_0^+ \bigl(x_a(t);x_b(t') \bigr) = \frac {-1} {4 \pi^2 \bigl[ (t - t' - i\epsilon) ^2 - L^2 \bigr]}\;,
\end{align}
and the detector response function (used in $A$) is obtained by letting $L \to 0$ and is also obtainable as the limit of Eq.~\eqref{eq:Ddetect} as $T \to 0$.  Both $X$ and $A$ can be evaluated analytically:
\begin{align}
\label{eq:X0}
	X_0 &= -\frac{e^{-\frac{L^2}{4 \sigma ^2}-\sigma ^2 \Omega ^2} \sigma \erfi \left(\frac{L}{2 \sigma }\right)} {4 L \sqrt{\pi }}\;, \\
\label{eq:A0}
	A_0 &= \frac{e^{-\sigma ^2 \Omega ^2}-\sqrt{\pi } \sigma \Omega \erfc (\sigma  \Omega )}{4 \pi }\;,
\end{align}
where $\sigma$ is the width of the window function (the time for which the detector is turned on), and the subscripts indicate that these are the Minkowski vacuum results, with $\erfi(z) = -i \erf(iz)$ and $\erfc(z) = 1-\erf(z)$, where $\erf(z)$ is the error function.  In the Minkowski vacuum case, the detectors become entangled if and only if $\abs {X_0} > A_0$.  This region in the $L$-$\Omega$ plane is above the slanted black line in Fig.~\ref{fig:threshold}.

Let's see what happens with a nonzero temperature.  Since we are interested in the possibility that the perceived thermality is due to de~Sitter expansion, we have a restriction on the temperature, which sets the scale for the cosmic horizon~$L_H = \kappa^{-1} = (2\pi T)^{-1}$.  If observers are to exist at all, this horizon must be much larger than their typical scale of experience, which can't be much smaller than~$\sigma$ if the detector is to be useful to them.  (Consider how useful a ``detector'' that operates on the scale of the Hubble time would be for humans.)  Thus, for de~Sitter expansion even to be a possibility, we require that $T \ll \sigma^{-1}$.

In both cases, the detector response function is given by Eq.~\eqref{eq:Ddetect}, while the Wightman function to be used in $X$ in the thermal case is~\cite{Weldon2000}
\begin{multline}
\label{eq:Dth}
	\Dth \bigl(x_a(t);x_b(t') \bigr) = \frac {T} {8 \pi L} \\ \times \Bigl\{ \coth \bigl[ \pi  T (L-y) \bigr]
 + \coth \bigl[ \pi T (L+y) \bigr] \Bigr\}
\end{multline}
and in the de~Sitter case is~\cite{Birrell1982}
\begin{multline}
\label{eq:DdS}
	\DdS \bigl(x_a(t);x_b(t') \bigr) = \\
		\left(\frac{-1}{4\pi ^2}\right)\left[\frac{\sinh^2(\pi  T y)}{\pi ^2 T^2}-e^{2 \pi  T x} L^2\right]^{-1}\;,
\end{multline}
where $x = t + t'$, and $y = t - t' - i\epsilon$ in both.  One can verify that in both cases, taking $L \to 0$ gives Eq.~\eqref{eq:Ddetect}, and taking $T \to 0$ gives Eq.~\eqref{eq:DMink}.

In both the thermal and de~Sitter cases, the integral in Eq.~\eqref{eq:Xdef} can be well approximated by an asymptotic series in~$T$ (as~$T \to 0$), generated from the Taylor expansion of $\Dth$ and $\DdS$, respectively, about $x=y=0$.  Although the radius of convergence of the Taylor series is finite, for any reasonable detector setup, we are requiring that $L \gg \sigma$.  Since the nearest pole is either $O(L)$ or $O(T^{-1})$ away, the Gaussian window function, whose width is much smaller than either $L$ or $T^{-1}$, will regularize, within the integral, any reasonably truncated Taylor approximation to the Wightman function.  This results in a valid asymptotic series for $X$ in either case, as $T \to 0$.  The integral in Eq.~\eqref{eq:Adef} can be done similarly by writing $D_T^+ = D_0^+ + \Delta D_T^+$ (noting that the pole at $y=0$ has been eliminated in $\Delta D_T^+$) and calculating the temperature-dependent correction to Eq.~\eqref{eq:A0}.  Numerical checks of particular cases verify that these approximations are valid.  The results are presented in Fig.~\ref{fig:threshold}.

Several points are in order here.  First, detectors see anything at all in the Minkowski vacuum case because the time-energy uncertainty relation, $\Delta t \Delta E \gtrsim \tfrac 1 2$, implies that a detector operating for a finite time has a nonzero probability $A_0$ of becoming excited, even when the field is in the vacuum state.  Entanglement exists when virtual particle exchange dominates over local noise.  When the magnitude of the exchange amplitude~$\abs {X_0}$ exceeds $A_0$, the detectors become entangled~\cite{Reznik2005,Horodecki1996}.  Because of how both functions scale with $\Omega$ and $L$, in the vacuum case one can always reduce the local noise below $\abs{X_0}$ by sufficiently increasing~$\Omega$.  In the thermal and de~Sitter cases, the local noise profile~$A$ fails to decrease fast enough for large~$\Omega$, resulting in a maximum entangling frequency for a given~$L$, as well as a maximum separation beyond which entanglement is impossible, regardless of~$\Omega$.

\begin{figure}[htbp]
\begin{center}
\includegraphics[width= 8.5 cm,angle=0]{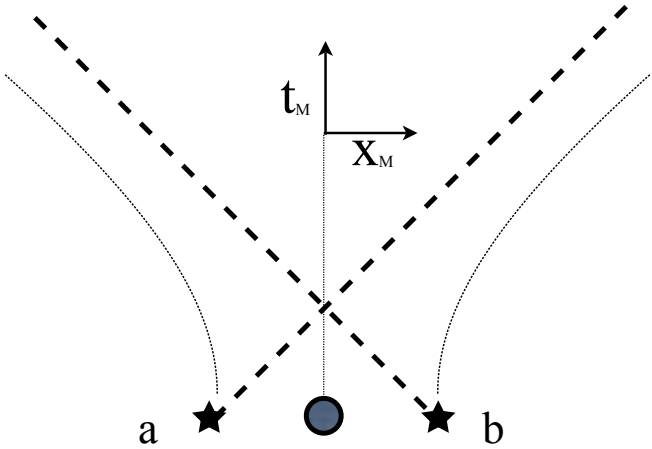}
\caption{Spacetime diagram in Minkowski coordinates in the rest frame of the home planet (circle). Null rays travel at 45 degrees and light dotted lines represent geodesics in de Sitter space. Messages sent from detectors $a$ or $b$ ($\star$) never reach the other detector because of the Hubble expansion of the universe. However, the home planet can receive and analyze the messages, differentiating the entanglement scenarios depicted in Fig.~\ref{fig:threshold} .}
\label{fig:spacetime}
\end{center}
\end{figure}

What does this mean for our curious planetary inhabitants?  Let's assume they have two satellites, with detectors of the sort we've been using, located on comoving trajectories as described above, with $\kappa^{-1} < L < 2\kappa^{-1} $ so that in the de~Sitter case they would be outside of each other's cosmic horizon but within that of the home planet (so they can still send messages to it, as described in Fig.~\ref{fig:spacetime}).  The satellites are programmed to interact the field locally with qubits having a resonant frequency that will lead to entanglement in the thermal case and to a separable state in the de~Sitter case (e.g., the red star in Fig.~\ref{fig:threshold}).  After the interactions, they each run a local measurement protocol that implements one side of a test of Bell inequality violation, after which they send data back to the home planet for analysis.  If thermality is a result of expansion, there will be no entanglement, but if it is a result of heating in flat spacetime, then the entanglement can be verified upon receipt of the transmissions from both satellites.  Because this effect only manifests when the detectors pass beyond each others' cosmic horizons (in the de~Sitter case), a third party is required to make the determination.

We have demonstrated that while expansion and heating give rise to the same (thermal) signature in a single inertial particle detector, for certain choices of detector parameters, a heated field in flat spacetime is able to entangle detector pairs that the conformal vacuum in the associated de~Sitter universe cannot.  Thus, the universes can be distinguished by their {\it entangling power}.  Two detectors are required and must be beyond each others' cosmic horizons (in the de~Sitter case) to see the effect.  Although, if present, the entanglement is exceedingly small, in principle its presence can always be determined by classical communication of local measurement data to a third party, as long as the verifier is able to receive messages from both detectors.  These results are contrary to the intuition that ``curvature generates entanglement'' between field modes~\cite{Ball2006}, since from it one would expect a {\it larger} entangled region in the de~Sitter case.  The ability of the field to swap its entanglement to local detectors is an operational question, though, and for this setup, the vacuum in a curved spacetime has less entangling power than a corresponding heated field in flat spacetime, even though both produce the same local detector response.


We thank John Preskill, Sean Carroll, Gerard Milburn, and Carl Caves for invaluable discussions, comments, and guidance.  N.C.M. thanks the faculty and staff of the Caltech Institute for Quantum Information for their hospitality during his visits, which allowed this work to come to fruition.  Both G.V.S. and N.C.M. acknowledge support from the National Science Foundation, with N.C.M. also supported by the U.S. Department of Defense.


\bibliography{CurveEnt}

\begin{thebibliography}{28}
\expandafter\ifx\csname natexlab\endcsname\relax\def\natexlab#1{#1}\fi
\expandafter\ifx\csname bibnamefont\endcsname\relax
  \def\bibnamefont#1{#1}\fi
\expandafter\ifx\csname bibfnamefont\endcsname\relax
  \def\bibfnamefont#1{#1}\fi
\expandafter\ifx\csname citenamefont\endcsname\relax
  \def\citenamefont#1{#1}\fi
\expandafter\ifx\csname url\endcsname\relax
  \def\url#1{\texttt{#1}}\fi
\expandafter\ifx\csname urlprefix\endcsname\relax\def\urlprefix{URL }\fi
\providecommand{\bibinfo}[2]{#2}
\providecommand{\eprint}[2][]{\url{#2}}

\bibitem[{\citenamefont{Bekenstein}(1973)}]{Bekenstein1973}
\bibinfo{author}{\bibfnamefont{J.~D.} \bibnamefont{Bekenstein}},
  \bibinfo{journal}{Phys. Rev. D} \textbf{\bibinfo{volume}{7}},
  \bibinfo{pages}{2333} (\bibinfo{year}{1973}).

\bibitem[{\citenamefont{Hawking}(1975)}]{Hawking1975}
\bibinfo{author}{\bibfnamefont{S.~W.} \bibnamefont{Hawking}},
  \bibinfo{journal}{Comm. Math. Phys.}
  \textbf{\bibinfo{volume}{43}}, \bibinfo{pages}{199} (\bibinfo{year}{1975}).

\bibitem[{\citenamefont{Gibbons and Hawking}(1977)}]{Gibbons1977}
\bibinfo{author}{\bibfnamefont{G.~W.} \bibnamefont{Gibbons}} \bibnamefont{and}
  \bibinfo{author}{\bibfnamefont{S.~W.} \bibnamefont{Hawking}},
  \bibinfo{journal}{Phys. Rev. D} \textbf{\bibinfo{volume}{15}},
  \bibinfo{pages}{2738} (\bibinfo{year}{1977}).

\bibitem[{\citenamefont{Bousso}(2002{\natexlab{a}})}]{Bousso2002a}
\bibinfo{author}{\bibfnamefont{R.}~\bibnamefont{Bousso}},
  \bibinfo{journal}{Rev. Mod. Phys.} \textbf{\bibinfo{volume}{74}},
  \bibinfo{pages}{825} (\bibinfo{year}{2002}{\natexlab{a}}).

\bibitem[{\citenamefont{Hawking et~al.}(2001)\citenamefont{Hawking, Maldacena,
  and Strominger}}]{Hawking2001}
\bibinfo{author}{\bibfnamefont{S.}~\bibnamefont{Hawking}},
  \bibinfo{author}{\bibfnamefont{J.}~\bibnamefont{Maldacena}},
  \bibnamefont{and}
  \bibinfo{author}{\bibfnamefont{A.}~\bibnamefont{Strominger}},
  \bibinfo{journal}{J. High Energy Phys.} \textbf{\bibinfo{volume}{0105}}, \bibinfo{pages}{001}
  (\bibinfo{year}{2001}).

\bibitem[{\citenamefont{Goheer et~al.}(2003)\citenamefont{Goheer, Kleban, and
  Susskind}}]{Goheer2003}
\bibinfo{author}{\bibfnamefont{N.}~\bibnamefont{Goheer}},
  \bibinfo{author}{\bibfnamefont{M.}~\bibnamefont{Kleban}}, \bibnamefont{and}
  \bibinfo{author}{\bibfnamefont{L.}~\bibnamefont{Susskind}},
  \bibinfo{journal}{J. High Energy Phys.} \textbf{\bibinfo{volume}{0307}}, \bibinfo{pages}{056}
  (\bibinfo{year}{2003}).

\bibitem[{\citenamefont{Bousso}(2002{\natexlab{b}})}]{Bousso2002}
\bibinfo{author}{\bibfnamefont{R.}~\bibnamefont{Bousso}}
  (\bibinfo{year}{2002}{\natexlab{b}}), \eprint{arXiv:hep-th/0205177}.

\bibitem[{\citenamefont{Ball et~al.}(2006)\citenamefont{Ball, Fuentes-Schuller,
  and Schuller}}]{Ball2006}
\bibinfo{author}{\bibfnamefont{J.~L.} \bibnamefont{Ball}},
  \bibinfo{author}{\bibfnamefont{I.}~\bibnamefont{Fuentes-Schuller}},
  \bibnamefont{and} \bibinfo{author}{\bibfnamefont{F.~P.}
  \bibnamefont{Schuller}}, \bibinfo{journal}{Phys. Lett. A}
  \textbf{\bibinfo{volume}{359}}, \bibinfo{pages}{550} (\bibinfo{year}{2006}).

\bibitem[{\citenamefont{Summers and Werner}(1987)}]{Summers1987}
\bibinfo{author}{\bibfnamefont{S.~J.} \bibnamefont{Summers}} \bibnamefont{and}
  \bibinfo{author}{\bibfnamefont{R.}~\bibnamefont{Werner}},
  \bibinfo{journal}{Comm. Math. Phys.}
  \textbf{\bibinfo{volume}{110}}, \bibinfo{pages}{247} (\bibinfo{year}{1987}).

\bibitem[{\citenamefont{Birrell and Davies}(1982)}]{Birrell1982}
\bibinfo{author}{\bibfnamefont{N.~D.} \bibnamefont{Birrell}} \bibnamefont{and}
  \bibinfo{author}{\bibfnamefont{P.~C.~W.} \bibnamefont{Davies}},
  \emph{\bibinfo{title}{Quantum Field Thoery in Curved Space}}
  (\bibinfo{publisher}{Cambridge}, \bibinfo{year}{1982}).

\bibitem[{\citenamefont{Nielsen and Chuang}(2000)}]{Nielsen2000}
\bibinfo{author}{\bibfnamefont{M.~A.} \bibnamefont{Nielsen}} \bibnamefont{and}
  \bibinfo{author}{\bibfnamefont{I.~L.} \bibnamefont{Chuang}},
  \emph{\bibinfo{title}{Quantum Computation and Quantum Information}}
  (\bibinfo{publisher}{Cambridge}, \bibinfo{year}{2000}).

\bibitem[{\citenamefont{Bennett et~al.}(1993)\citenamefont{Bennett, Brassard,
  Cr\'epeau, Jozsa, Peres, and Wootters}}]{Bennett1993}
\bibinfo{author}{\bibfnamefont{C.~H.} \bibnamefont{Bennett}},
  \bibinfo{author}{\bibfnamefont{G.}~\bibnamefont{Brassard}},
  \bibinfo{author}{\bibfnamefont{C.}~\bibnamefont{Cr\'epeau}},
  \bibinfo{author}{\bibfnamefont{R.}~\bibnamefont{Jozsa}},
  \bibinfo{author}{\bibfnamefont{A.}~\bibnamefont{Peres}}, \bibnamefont{and}
  \bibinfo{author}{\bibfnamefont{W.~K.} \bibnamefont{Wootters}},
  \bibinfo{journal}{Phys. Rev. Lett.} \textbf{\bibinfo{volume}{70}},
  \bibinfo{pages}{1895} (\bibinfo{year}{1993}).

\bibitem[{\citenamefont{Lo and Lutkenhaus}(2007)}]{Lo2007}
\bibinfo{author}{\bibfnamefont{H.-K.} \bibnamefont{Lo}} \bibnamefont{and}
  \bibinfo{author}{\bibfnamefont{N.}~\bibnamefont{Lutkenhaus}}
  (\bibinfo{year}{2007}), \eprint{arXiv:quant-ph/0702202}.

\bibitem[{\citenamefont{Masanes et~al.}(2007)\citenamefont{Masanes, Liang, and
  Doherty}}]{Masanes2007}
\bibinfo{author}{\bibfnamefont{L.}~\bibnamefont{Masanes}},
  \bibinfo{author}{\bibfnamefont{Y.-C.} \bibnamefont{Liang}}, \bibnamefont{and}
  \bibinfo{author}{\bibfnamefont{A.~C.} \bibnamefont{Doherty}}
  (\bibinfo{year}{2007}), \eprint{arXiv:quant-ph/0703268}.

\bibitem[{\citenamefont{Blume-Kohout et~al.}(2002)\citenamefont{Blume-Kohout,
  Caves, and Deutsch}}]{Blume-Kohout2002}
\bibinfo{author}{\bibfnamefont{R.}~\bibnamefont{Blume-Kohout}},
  \bibinfo{author}{\bibfnamefont{C.}~\bibnamefont{Caves}}, \bibnamefont{and}
  \bibinfo{author}{\bibfnamefont{I.}~\bibnamefont{Deutsch}},
  \bibinfo{journal}{Found. Phys.} \textbf{\bibinfo{volume}{32}},
  \bibinfo{pages}{1641} (\bibinfo{year}{2002}).

\bibitem[{\citenamefont{Reznik et~al.}(2005)\citenamefont{Reznik, Retzker, and
  Silman}}]{Reznik2005}
\bibinfo{author}{\bibfnamefont{B.}~\bibnamefont{Reznik}},
  \bibinfo{author}{\bibfnamefont{A.}~\bibnamefont{Retzker}}, \bibnamefont{and}
  \bibinfo{author}{\bibfnamefont{J.}~\bibnamefont{Silman}},
  \bibinfo{journal}{Phys. Rev. A}
  \textbf{\bibinfo{volume}{71}}, \bibinfo{eid}{042104}
  (\bibinfo{year}{2005}).

\bibitem[{\citenamefont{Massar and Spindel}(2006)}]{Massar2006}
\bibinfo{author}{\bibfnamefont{S.}~\bibnamefont{Massar}} \bibnamefont{and}
  \bibinfo{author}{\bibfnamefont{P.}~\bibnamefont{Spindel}},
  \bibinfo{journal}{Phys. Rev. D} \textbf{\bibinfo{volume}{74}}, \bibinfo{eid}{085031}
  (\bibinfo{year}{2006}).

\bibitem[{\citenamefont{Braun}(2005)}]{Braun2005}
\bibinfo{author}{\bibfnamefont{D.}~\bibnamefont{Braun}},
  \bibinfo{journal}{Phys. Rev. A}
  \textbf{\bibinfo{volume}{72}}, \bibinfo{eid}{062324}
  (\bibinfo{year}{2005}).

\bibitem[{\citenamefont{Spradlin et~al.}(2001)\citenamefont{Spradlin,
  Strominger, and Volovich}}]{Spradlin2001}
\bibinfo{author}{\bibfnamefont{M.}~\bibnamefont{Spradlin}},
  \bibinfo{author}{\bibfnamefont{A.}~\bibnamefont{Strominger}},
  \bibnamefont{and} \bibinfo{author}{\bibfnamefont{A.}~\bibnamefont{Volovich}}
  (\bibinfo{year}{2001}), \eprint{arXiv:hep-th/0110007}.

\bibitem[{\citenamefont{Guth}(1997)}]{Guth1997}
\bibinfo{author}{\bibfnamefont{A.~H.} \bibnamefont{Guth}},
  \emph{\bibinfo{title}{The Inflationary Universe: The Quest for a New Theory
  of Cosmic Origins}} (\bibinfo{publisher}{Basic Books}, \bibinfo{year}{1997}).

\bibitem[{\citenamefont{DeWitt}(1979)}]{DeWitt1979}
\bibinfo{author}{\bibfnamefont{B.~S.} \bibnamefont{DeWitt}}, in
  \emph{\bibinfo{booktitle}{General Relativity, An Einstein Centenary Survey}}
  (\bibinfo{publisher}{Cambridge}, \bibinfo{year}{1979}).

\bibitem[{\citenamefont{Masanes}(2006)}]{Masanes2006}
\bibinfo{author}{\bibfnamefont{L.}~\bibnamefont{Masanes}},
  \bibinfo{journal}{Phys. Rev. Lett.} \textbf{\bibinfo{volume}{97}},
  \bibinfo{eid}{050503} (\bibinfo{year}{2006}).

\bibitem[{\citenamefont{Sriramkumar and Padmanabhan}(1996)}]{Sriramkumar1996}
\bibinfo{author}{\bibfnamefont{L.}~\bibnamefont{Sriramkumar}} \bibnamefont{and}
  \bibinfo{author}{\bibfnamefont{T.}~\bibnamefont{Padmanabhan}},
  \bibinfo{journal}{Class. Quant. Grav.}
  \textbf{\bibinfo{volume}{13}}, \bibinfo{pages}{2061} (\bibinfo{year}{1996}).

\bibitem[{\citenamefont{Vidal and Werner}(2002)}]{Vidal2002}
\bibinfo{author}{\bibfnamefont{G.}~\bibnamefont{Vidal}} \bibnamefont{and}
  \bibinfo{author}{\bibfnamefont{R.~F.} \bibnamefont{Werner}},
  \bibinfo{journal}{Phys. Rev. A} \textbf{\bibinfo{volume}{65}},
  \bibinfo{pages}{032314} (\bibinfo{year}{2002}).

\bibitem[{\citenamefont{Horodecki et~al.}(1996)\citenamefont{Horodecki,
  Horodecki, and Horodecki}}]{Horodecki1996}
\bibinfo{author}{\bibfnamefont{M.}~\bibnamefont{Horodecki}},
  \bibinfo{author}{\bibfnamefont{P.}~\bibnamefont{Horodecki}},
  \bibnamefont{and}
  \bibinfo{author}{\bibfnamefont{R.}~\bibnamefont{Horodecki}},
  \bibinfo{journal}{Phys. Lett. A} \textbf{\bibinfo{volume}{223}},
  \bibinfo{pages}{1} (\bibinfo{year}{1996}).

\bibitem[{\citenamefont{Horodecki et~al.}(1997)\citenamefont{Horodecki,
  Horodecki, and Horodecki}}]{Horodecki1997}
\bibinfo{author}{\bibfnamefont{M.}~\bibnamefont{Horodecki}},
  \bibinfo{author}{\bibfnamefont{P.}~\bibnamefont{Horodecki}},
  \bibnamefont{and}
  \bibinfo{author}{\bibfnamefont{R.}~\bibnamefont{Horodecki}},
  \bibinfo{journal}{Phys. Rev. Lett.} \textbf{\bibinfo{volume}{78}},
  \bibinfo{pages}{574} (\bibinfo{year}{1997}).

\bibitem[{\citenamefont{Peskin and Schroeder}(1995)}]{Peskin1995}
\bibinfo{author}{\bibfnamefont{M.~E.} \bibnamefont{Peskin}} \bibnamefont{and}
  \bibinfo{author}{\bibfnamefont{D.~V.} \bibnamefont{Schroeder}},
  \emph{\bibinfo{title}{An Introduction to Quantum Field Theory}}
  (\bibinfo{publisher}{Perseus}, \bibinfo{year}{1995}).

\bibitem[{\citenamefont{Weldon}(2000)}]{Weldon2000}
\bibinfo{author}{\bibfnamefont{H.~A.} \bibnamefont{Weldon}},
  \bibinfo{journal}{Phys. Rev. D} \textbf{\bibinfo{volume}{62}},
  \bibinfo{pages}{056010} (\bibinfo{year}{2000}).

\end{thebibliography}
\end{document}